\documentclass[aps,reprint,superscriptaddress]{revtex4-1}

\usepackage{gensymb}
\usepackage{graphicx} 
\usepackage{amsmath} 
\usepackage{placeins}
\usepackage{booktabs}
\usepackage{multirow}
\usepackage{caption}
\usepackage{subcaption}
\usepackage{textcomp}
\usepackage{hyperref}

\begin{document}

\title{Dipolar-coupled moment correlations in clusters of magnetic nanoparticles}

\author{P. Bender}
\email{philipp.bender@uni.lu, current address: University of Luxembourg, 1511 Luxembourg, Grand Duchy of Luxembourg, Luxembourg.}
\affiliation{Departamento CITIMAC, Faculty of Science, University of Cantabria, 39005 Santander, Spain}

\author{E. Wetterskog}
\affiliation{Department of Engineering Sciences, Uppsala University, 75105 Uppsala, Sweden}

\author{D. Honecker}
\affiliation{Institut Laue-Langevin, 38042 Grenoble, France}

\author{J. Fock}
\affiliation{Technical University of Denmark, 2800 Kongens Lyngby, Denmark}

\author{C. Frandsen}
\affiliation{Technical University of Denmark, 2800 Kongens Lyngby, Denmark}

\author{C. Moerland}
\affiliation{Department of Applied Physics, Technische Universiteit Eindhoven, Eindhoven, The
Netherlands}

\author{L.~K. Bogart}
\affiliation{UCL Healthcare Biomagnetics Laboratory, University College London, 21 Albemarle Street, London, W1S 4BS, UK}

\author{O. Posth}
\affiliation{Physikalisch-Technische Bundesanstalt, 10587 Berlin, Germany}

\author{W. Szczerba}
\affiliation{Bundesanstalt f\"{u}r Materialforschung und -pr\"{u}fung, 12205 Berlin, Germany.}
\affiliation{AGH University of Science and Technology, 30-059 Krakow, Poland.}

\author{H. Gavil\'{a}n}
\affiliation{Instituto de Ciencia de Materiales de Madrid, ICMM/CSIC, Madrid, Spain}

\author{R. Costo}
\affiliation{Instituto de Ciencia de Materiales de Madrid, ICMM/CSIC, Madrid, Spain}

\author{M.~T. Fern\'andez-D\'iaz}
\affiliation{Institut Laue-Langevin, 38042 Grenoble, France}

\author{D. Gonz\'{a}lez-Alonso}
\affiliation{Departamento CITIMAC, Faculty of Science, University of Cantabria, 39005 Santander, Spain}

\author{L. Fern\'{a}ndez Barqu\'{i}n}
\affiliation{Departamento CITIMAC, Faculty of Science, University of Cantabria, 39005 Santander, Spain}

\author{C. Johansson}
\affiliation{RISE Acreo, 40014 G\"{o}teborg, Sweden}

\date{\today}

\begin{abstract}
Here, we resolve the nature of the moment coupling between 10-nm DMSA-coated magnetic nanoparticles. 
The individual iron oxide cores were composed of $>95\,\%$ maghemite and agglomerated to clusters.
At room temperature the ensemble behaved as a superparamagnet according to M\"ossbauer and magnetization measurements, however, with clear signs of dipolar interactions.
Analysis of temperature-dependent AC susceptibility data in the superparamagnetic regime indicates a tendency for dipolar coupled anticorrelations of the core moments within the clusters.
To resolve the directional correlations between the particle moments we performed polarized small-angle neutron scattering and determined the magnetic spin-flip cross-section of the powder in low magnetic field at 300\,K.
We extract the underlying magnetic correlation function of the magnetization vector field by an indirect Fourier transform of the cross-section.
The correlation function suggests non-stochastic preferential alignment between neighboring moments despite thermal fluctuations, with anticorrelations clearly dominating for next-nearest moments.
These tendencies are confirmed by Monte Carlo simulations of such core-clusters.

\end{abstract}


\maketitle

\section{Introduction}
Understanding the precise influence of dipolar interactions on the magnetization behavior of magnetic nanoparticles is of utmost importance for potential technological or biomedical applications \cite{nie2010properties,majetich2006magnetostatic}.
In particular the magnetic hyperthermia performance of nanoparticle ensembles can significantly depend on the core arrangement \cite{serantes2010influence,sadat2014effect,salas2014modulation,tan2014magnetic,sanz2016silico,ovejero2016effects,anand2016spin,usov2017interaction,bender2018influence}, mainly due to the sensitivity of magnetic relaxation on induced dipolar interaction energy \cite{morozov1987magnetic,iglesias2004magnetic,morup2010magnetic,dejardin2011magnetic,landi2014role,hovorka2014role,ivanov2016revealing,ivanov2017influence,dejardin2017effect}.
Experimentally, interacting nanoparticle ensembles have been much characterized \textit{via} temperature-dependent magnetometry techniques \cite{dormann1988dynamic,djurberg1997dynamics,jonsson1998dynamic,hansen2002critical,binns2002magnetic,poddar2002dipolar,andersson2015size}.
Regarding the blocking temperature $T_\mathrm{B}$, it was found in some studies that dipolar interactions increase $T_\mathrm{B}$ \cite{dormann1988dynamic,vargas2005effect}, whereas in other cases a reduction of $T_\mathrm{B}$ was observed \cite{el1992susceptibility,morup1994superparamagnetic}.
Corresponding simulations of randomly arranged 3D systems show that, on the nanoscale, this collective behavior is usually accompanied by a mixture of short-range ferromagnetic-(FM)-like and antiferromagnetic-(AFM)-like ordering of neighboring core moments for temperatures lower than $T_\mathrm{B}$ \cite{chantrell2000calculations,panissod2003magnetic}.

In low-dimensional (1-2D) systems, electron holography studies have revealed both nearest neighbor and overall FM-like ordering \cite{varon2013dipolar} in long and narrow ensembles in both close-packed as well as more disordered nanoparticle ensembles.
In thicker nanoparticle structures, long-range AFM-like interactions become important, as evidenced by super-spin domain formation, with sharp 180 degree walls between nearest neighbor cores \cite{varon2015longitudinal,jordanovic2015simulations}.
Experimental evidence for nearest neighbor moment correlations within ordered 3D arrays of magnetic cores was obtained by \citet{faure20132d}, using dynamic magnetometry in combination with Monte-Carlo simulations. 
The authors observed an increased tendency for the transition of a FM-like to an AFM-like moment order with increasing film thickness.
A special class of 3D ensembles of magnetic nanoparticles are particle core aggregates or clusters, also referred to as multi-core nanoparticles \cite{ijms160920308}.
Investigation of such particles has attracted much interest in recent years \cite{ahrentorp2015effective,ilg2017equilibrium,ludwig2014magnetic,bender2017structural}, mainly motivated by their large potential for biomedical applications \cite{blanco2015high,gutierrez2015synthesis}.

It is often assumed that interacting magnetic cores in the superparamagnetic regime, in which each individual core has a thermally fluctuating moment, do not display a self sustained ordered state \cite{allia2001granular,allia2011dynamic,morup2010magnetic}.
Yet, recent numerical calculations of core-clusters indicate that directional correlations may be indeed possible \cite{hovorka2014role}.

Experimental evidence for a correlated, thermally activated motion was obtained by resonant magnetic X-ray scattering for densely packed Co \cite{kortright2005interparticle} and iron oxide nanoparticles \cite{chesnel2018unraveling}.
Few other techniques, however, are sensitive to moment correlations on the interparticle length scale and even less are simultaneously capable of taking a \textit{snapshot} of the internal moment structure during superparamagnetic relaxation.
As a consequence, experimental observations of dipolar coupling in the superparamagnetic regime are still severely lacking. 

In this work, we exploit polarized small-angle neutron scattering (SANS) to gain information about directional correlations between the moments within clusters of 10-nm iron oxide cores in the superparamagnetic regime, here at 300\,K. 
Elastic neutron scattering has a measurement time scale on the order of a picosecond \cite{felber1998coherence} and is therefore capable of capturing \textit{snapshots} of the magnetic ordering  within core-clusters, in which relaxations of the entire cluster occurs on longer time scales (ns regime).
Furthermore, SANS provides information about magnetic correlations on the nanoscale and offers thus a unique approach to study magnetic nanoparticle systems, as also done in other studies \cite{bellouard1996magnetic,sankar2000magnetic,avdeev2004magnetic,farrell2006small,sachan2008field,alba2016magnetic,orue2018configuration}.
We, however, performed a complete longitudinal neutron-spin analysis in SANS (POLARIS) \cite{honecker2010longitudinal}, through which we were able to detect the purely magnetic scattering cross sections.
By using a model-independent analysis, based on indirect Fourier transformations, we extracted the underlying magnetic correlation functions in order to obtain information about the nature of the moment correlations.
Additionally, we have performed Monte-Carlo simulations to support our observations.

\section{Methods}
Iron oxide cores were prepared by thermal decomposition of an iron oleate in 1-octadecene and transferred to water in a subsequent ligand exchange step using dimercaptosuccinic acid (DMSA) \cite{salas2012controlled}.
The DMSA coating around the individual cores provides an insulating separation and as such should limit exchange interactions between surface atoms of neighboring cores even in case of agglomeration.
To prepare the powder samples, the colloidal dispersion was freeze-dried in a LYOQUEST-55 ECO and afterwards slightly compressed.

Transmission electron microscopy (TEM) images of the cores were taken with a 100-keV JEOL-JEM 1010 microscope.
The sample was prepared by placing a drop of the colloidal dispersion onto a carbon coated copper grid and allowing it to dry at room temperature.
The core size distribution was determined by measuring the diameter of 300 cores using the public domain \textit{ImageJ} software \cite{schneider2012nih}.
The obtained histogram was fitted with a standard lognormal function. 

Small-angle X-ray scattering (SAXS) measurements of the colloidal dispersion at 300\,K were performed using a Kratky system with slit focus, SAXSess by Anton Paar.
The data were corrected from background scattering and deconvoluted with the beam profile using the implemented \textit{SAXS-Quant} software.

A dynamic light scattering (DLS) measurement was conducted with a Malvern Zetasizer Nano ZS. 
The autocorrelation function was recorded in the $173^\circ$ backscatter mode.
The data analysis was performed with the nonlinear-non-negative least square
(NNLS) method.

Neutron diffraction of the powder was conducted at 300\,K using the high-resolution powder diffractometer D2B at the Institut Laue Langevin \cite{ILLproposalND}.
The powder was loaded into a vanadium can and the diffraction pattern was measured within the $2\Theta$ range $\sim3-160^\circ$ in steps of $0.05^\circ$ with a wavelength of 1.594\,\AA, covering the $q$-range $\sim2.4-77.4\,\mathrm{nm^{-1}}$.
The pattern was adjusted by a Rietveld analysis using the \textit{Fullprof Suite} program \cite{rodriguez1993recent}.
To describe the peak profile a Thompson-Cox-Hastings function was selected, which ensures a good description of the width excess arising from the average crystal size ($d$) and microstrain ($\epsilon$) of the core.

$^{57}$Fe M\"ossbauer spectroscopy of the frozen dispersion and the powder was performed using a conventional constant acceleration spectrometer with a source of $^{57}$Co in rhodium. 
Calibration was carried out at room temperature using a 12.5 $\mu$m $\alpha$-Fe foil.
A closed helium refrigerator from APD Cryogenics was used to cool the sample. 
The spectra were folded and calibrated and the spectra fitted in \textit{MATLAB} (MathWorks Inc., USA) using a previously described protocol \cite{Fock2016}.

Room temperature (RT) M\"ossbauer measurements (295(5)\,K) were obtained using conventional spectrometer from SeeCo Inc (USA) which operated in the constant acceleration mode, in transmission geometry, with $^{57}$Co in Rh foil as the source of the 14.4\,keV $\gamma$-rays. 
Velocity calibration was performed by recording a reference spectrum from a $10\,\mathrm{\mu m}$ thick foil of $\alpha$-Fe at room temperature. 
Measured spectra were folded and baseline corrected using cubic spline parameters derived from fitting the $\alpha$-Fe calibration spectrum, following a protocol implemented in the \textit{Recoil} analysis program \cite{fock2017centre}.

Field-dependent DC magnetization ($M(H)$) and temperature-dependent AC susceptibility (ACS($T$)) curves of the colloidal dispersion as well as powder were measured using a Quantum design MPMS XL SQUID, equipped with the ultra low-field option.
The diamagnetic background signals of the sample holder (ceramic cylinder) and water (in case of the colloidal dispersion) were subtracted and the magnetic moment was normalized either to the volume of the magnetic material or the iron mass.

Polarized SANS was conducted at 300\,K at the instrument D33 \cite{Dewhurst:ks5488} at the Institut Laue Langevin \cite{ILLproposal}.
A mean wavelength of $\lambda=0.6\,\mathrm{nm}$ was used, with a spread of $\Delta\lambda/\lambda\approx 10\,\%$.
The detector was located a distance of 13.4 and 3\,m respectively, yielding a corresponding scattering vector ($q$) range of $0.07-0.77\,\mathrm{nm^{-1}}$.
By employing longitudinal neutron spin analysis (POLARIS) \cite{honecker2010longitudinal}, we were able to resolve all four neutron spin intensities $I^{++}(\textbf{q})$, $I^{--}(\textbf{q})$, $I^{+-}(\textbf{q})$, $I^{-+}(\textbf{q})$.
For further information see the \hyperref[appendix]{appendix}.
The spin-flip cross section will be denoted as $I^{\mathrm{sf}}(\textbf{q})=I^{+-}(\textbf{q})=I^{-+}(\textbf{q})$ in the text.
The spin-leakage correction was performed by using \textit{GRASP} \cite{GRASP}.
A homogeneous horizontal magnetic field $\mathbf{H}\parallel\mathbf{e}_z$ was applied at the sample position perpendicular to the wavevector $\mathbf{k}_0\parallel\mathbf{e}_x$ of the incident neutron beam.
A minimum field strength of $\mu_0H=2\,\mathrm{mT}$ was necessary to provide a sufficient guide field to maintain the polarization of the neutrons.

\section{Experimental results \& discussion}
Here, we perform a thorough structural and magnetic characterization of the particles before we present the analysis of the polarized SANS experiment and the Monte Carlo simulations.
The size and shape of the particles in colloidal dispersion is determined using TEM, SAXS and DLS.
Information regarding the chemical composition, crystallinity and magnetic structure is obtained using neutron diffraction and M\"ossbauer spectroscopy. 
For the former, the particles were in powder form whilst for the latter, measurements were obtained from both the (frozen) colloid as well the powder.  
Analysis of the M\"ossbauer spectra also provides information about the coupling between the core moments in the colloid and the powder.
The influence of dipolar interactions between the cores is probed using both DC and AC magnetometry, for particles in both colloidal and powder form. 

\subsection{Structural and magnetic pre-characterization}

\begin{figure}[t]
\centering
\includegraphics[width=1\columnwidth]{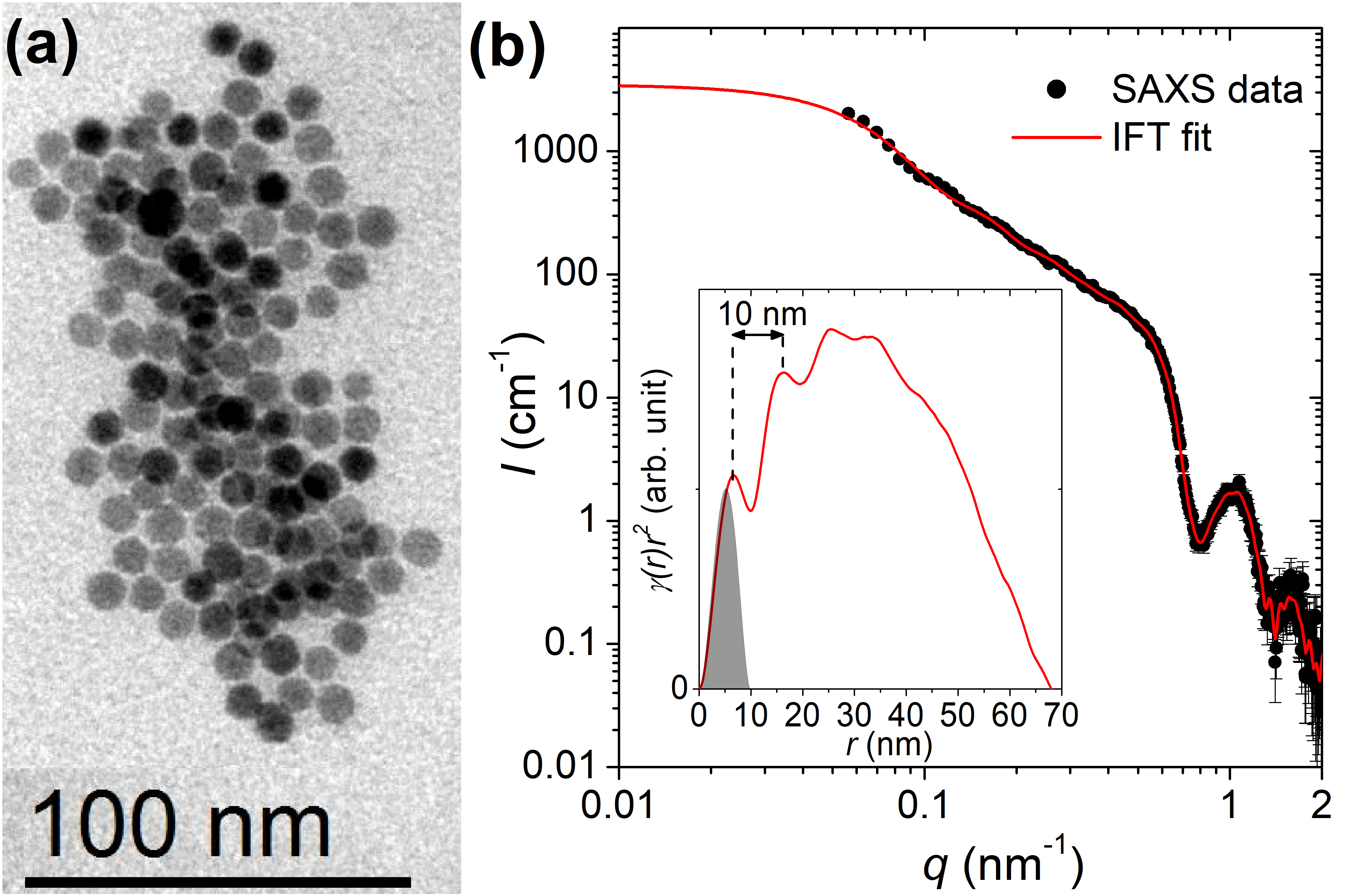}
\caption{\label{Fig1}(a) TEM image of the iron oxide cores. 
(b) Measured SAXS intensity $I(q)$ (radial average, $T=300\,\mathrm{K}$) of the dispersion and fit by IFT. \textit{inset:} Correlation function $C(r)r^2$ derived by the IFT of $I(q)$ and the expected profile of a homogeneous sphere calculated with Eq.\,\ref{Eq1} for $D=10\,\mathrm{nm}$ (grey area).}
\end{figure}

A representative TEM image is shown in Fig.\,\ref{Fig1}(a).
As can be seen, the cores are spherically shaped and nearly monodisperse.
The mean core size $\left\langle D_\mathrm{TEM}\right\rangle=9.7\,\mathrm{nm}$ and the polydispersity index very low with $\mathrm{PDI}=0.06$ (standard deviation/mean). 
Additionally it can be observed that the cores are separated from each other by around 1\,nm.
This can be attributed to the DMSA coating, which prevents a direct contact between the cores.

From the SAXS data we derived the underlying correlation function $C(r)r^2$ by an indirect Fourier transform (IFT) \cite{bender2017structural,bender2018relating,glatter1977new,hansen2000bayesian,Vestergaard:wf5022} of the radially averaged scattering intensity $I(q)$ (Fig.\,\ref{Fig1}(b)).
Here, $C(r)$ is the autocorrelation function of the nuclear scattering length density, which provides useful information about the 3D averaged spatial distribution of the particles in the colloid.
For individually dispersed spherical cores with radius $R$, one would expect 
\begin{equation}\label{Eq1}
C(r)r^2=3\left(\frac{r}{R}\right)^2\left(1-\frac{3r}{4R}+\frac{1}{16}\left(\frac{r}{R}\right)^3\right),
\end{equation}
which is a nearly bell-shaped profile.
Comparison of the determined correlation function for our sample reveals a significant deviation from the profile calculated for a 10-nm sphere (\textit{inset} of Fig.\,\ref{Fig1}(b)), and instead suggests that the cores were in fact agglomerated to clusters, probably induced by dipolar interactions \cite{pyanzina2017cluster}, with maximal lengths of around 68\,nm.

The presence of large agglomerates is confirmed by DLS, from which we obtained a $z$-average (i.e. mean intensity weighted hydrodynamic size) of 79\,nm.

\begin{figure}[b]
\centering
\includegraphics[width=1\columnwidth]{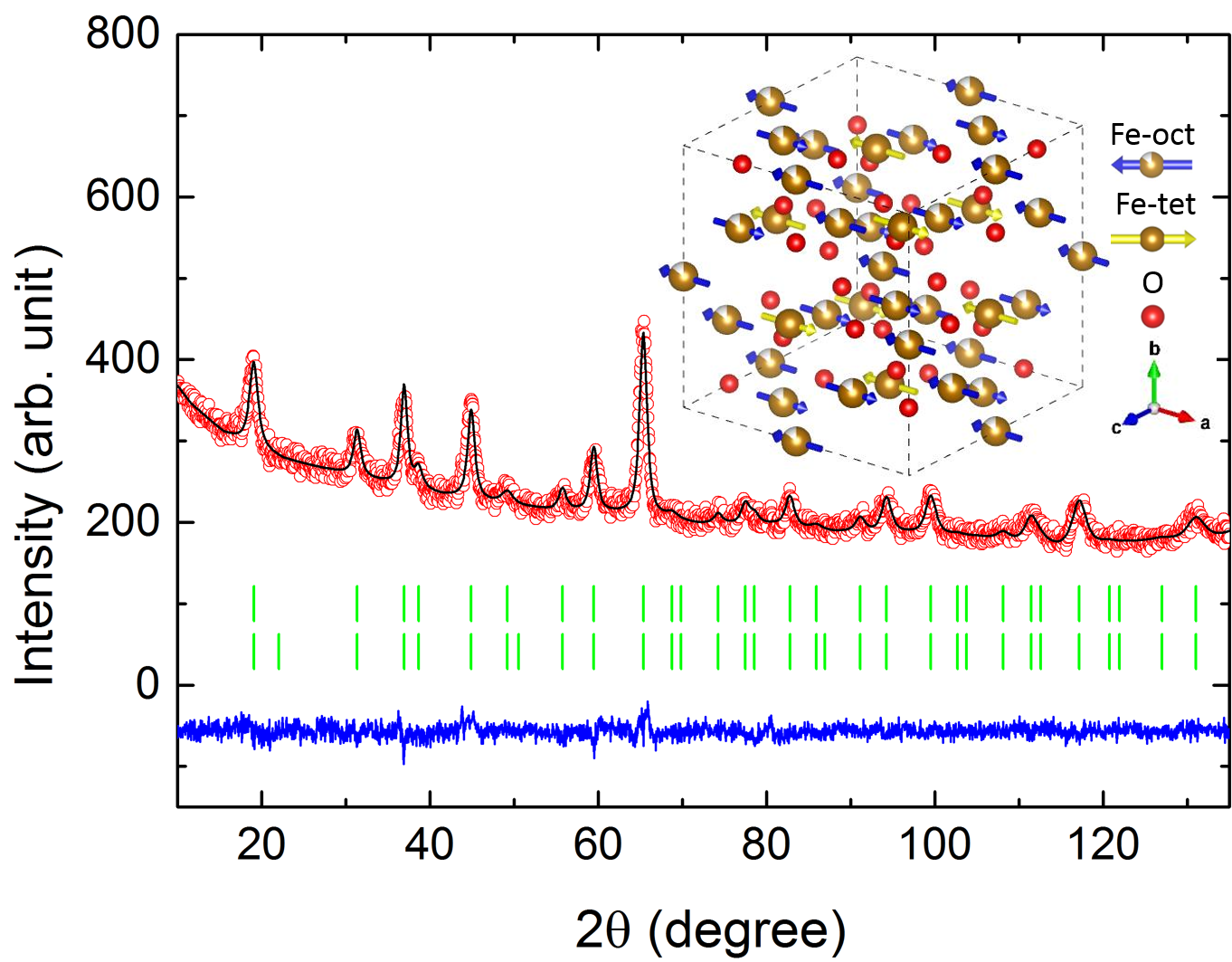}
\caption{\label{Fig2}Neutron diffraction pattern resolved by the Rietveld method. Residuals are represented by the blue line and the vertical tick marks indicate the positions of the nuclear (top) and magnetic (bottom) diffraction peaks. \textit{inset}: Ferrimagnetic structure of the iron oxide cores.}
\end{figure}

\begin{figure*}[ht]
\centering
\includegraphics[width=1\textwidth]{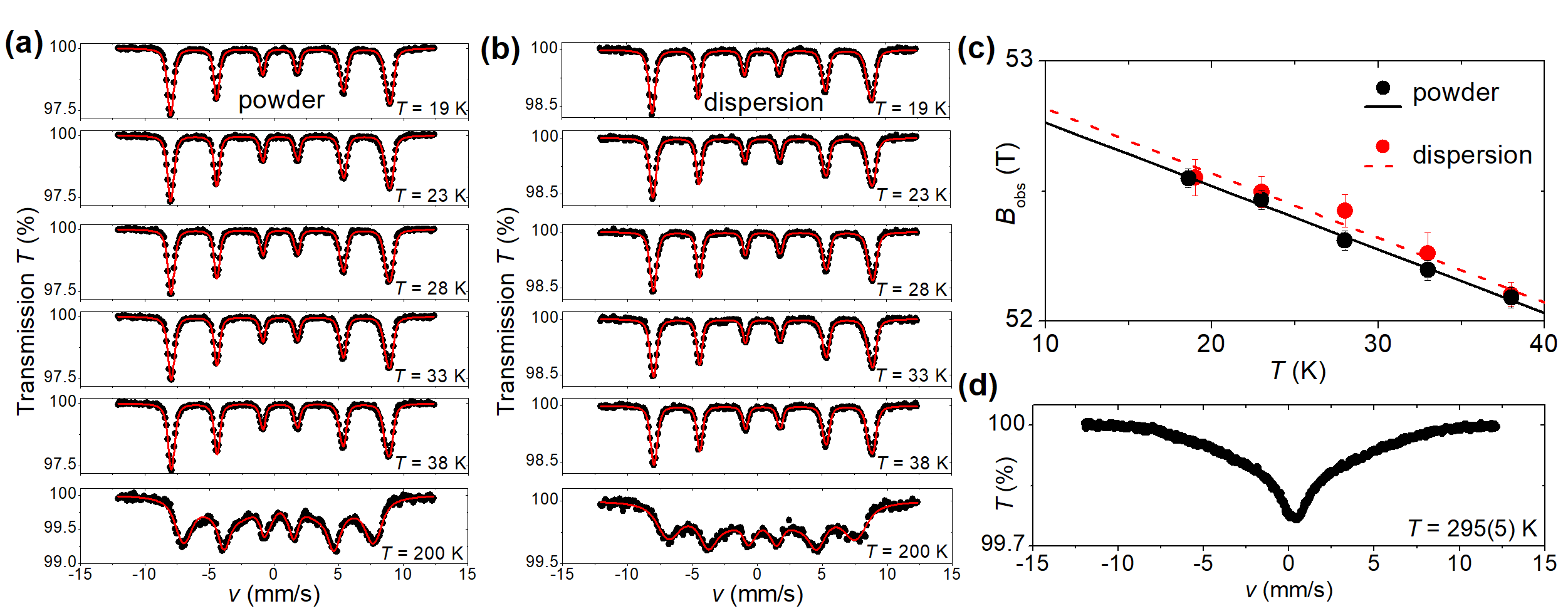}
\caption{\label{Fig3}M\"ossbauer spectra and associated fits of the powder sample (a) and the frozen dispersion (b) at the indicated temperatures.  
(c) Comparison of the mean hyperfine field  for the powder (black) and the frozen colloidal dispersion (red) for $T<40\,\mathrm{K}$.
Both lines are linear fits.
(d) RT M\"ossbauer spectrum of the particle powder.}
\end{figure*}

The Rietveld refined neutron diffraction pattern of the powder is shown in Fig.\,\ref{Fig2}. 
All reflections can be indexed by a cubic $Fd\bar{3}m$ space group with a lattice parameter $a = 8.3565(3)\,\mathrm{\AA}$.
This suggests that the iron oxide is dominated by the maghemite phase ($\gamma$-$\mathrm{Fe_2O_3}$, $a = 8.34\,\mathrm{\AA}$) \cite{shmakov1995vacancy,fock2017centre}, with minor presence of magnetite ($\mathrm{Fe_3O_4}$, $a = 8.39\,\mathrm{\AA}$) \cite{okudera1996temperature}.
The structural parameters refined at 300\,K of both nuclear and magnetic contributions are summarized in Table\,\ref{Table A} in the \hyperref[appendix]{appendix}. 
As expected for magnetite-maghemite mixtures, these measurements reveal ferrimagnetic ordering, with an average moment of $4.5(2)$ and $4.2(2)\,\mu_\mathrm{B}$ per Fe ion at the tetrahedral and octahedral sites of the inverse spinel, respectively \cite{shmakov1995vacancy,fock2017centre}. 
From this, we derive a net magnetic moment of $1.9(4)\,\mu_\mathrm{B}$ per formula unit and a volume saturation magnetization value of $330(60)\,\mathrm{kA/m}$, which is close to the one calculated from the DC magnetization and discussed in the following section.
The average crystal size $D_\mathrm{cryst}$ was determined to be 9(1)\,nm, and which agrees well with core size according to TEM and thus confirms that the cores are single-crystalline.
In comparison, the magnetic core size $D_\mathrm{mag}$ is reduced to 6(2)\,nm.
This discrepancy between crystal and magnetic core size indicates a surface layer of around 1.5\,nm of uncorrelated surface spins \cite{andersson2015size,Iglesias2004286,negi2017surface,de2017remanence,bender2017distribution}, as also observed by polarized SANS studies of similar systems \cite{krycka2010core,1367-2630-14-1-013025}.

\begin{figure*}[ht]
\centering
\includegraphics[width=1\textwidth]{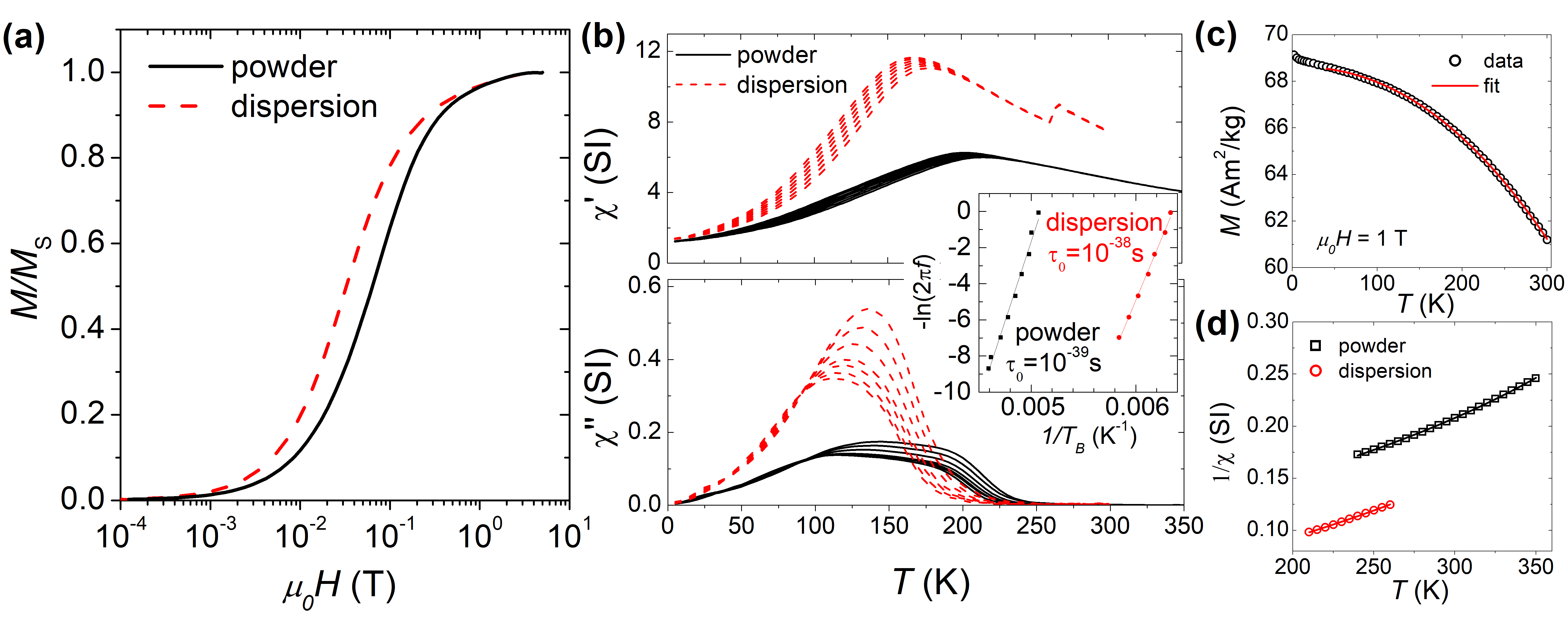}
\caption{\label{Fig4}
(a) Normalized isothermal magnetization curves $M(H)/M_\mathrm{S}$ of the core-clusters in powder form and colloidal dispersion measured at $T=300\,\mathrm{K}$.
(b) In-phase- ($\chi'$) and out-of-phase ($\chi''$) components of the ACS($T$) measurements of the dispersion and the powder for frequencies $f=0.17-170\,\mathrm{Hz}$ (from top to bottom). 
\textit{inset}: Fit of the frequency-dependency of the blocking temperatures $T_\mathrm{B} (f)$ to determine the characteristic attempt times $\tau_0$.
(c) High-field ($\mu_0H=1\,\mathrm{T}$) magnetization \textit{vs.} temperature of the powder. The line corresponds to a modified Bloch-law in the temperature range 40-300 K. The slight upturn at low T (excluded from the fit) corresponds primarily to paramagnetic impurities in the sample cup and straw.
(d) Inverse susceptibility ($1/\chi$) \textit{vs.} temperature of the freeze dried powder, and the frozen dispersion (here using $f = 0.17\,\mathrm{Hz}$). The solid lines corresponds to fits to a (Curie-Weiss) mean-field model (Eq.\,\ref{Eq2}).
The colloid thawed at around 260\,K, as indicated by the kink in $\chi'$ in panel (b), and thus the analysis was restricted to temperatures $T<260\,\mathrm{K}$.}
\end{figure*}

The temperature-dependent M\"ossbauer spectra for the particles in both (a) the powder and (b) the colloidal dispersion are compared in Fig.\,\ref{Fig3}.
The mean isomer shift, determined using the model independent method described by the authors Fock \& Bogart \textit{et al.} \cite{fock2017centre}, allows us to conclude that the cores have a composition of $95/5\,\mathrm{wt\%}$  $\gamma$-$\mathrm{Fe_2O_3}$/$\mathrm{Fe_3O_4}$, which is in excellent agreement with that observed in our neutron diffraction measurements.
Analysis of the temperature dependency of the M\"ossbauer spectra provides additional information about the nature of the interactions between cores. 
For temperatures well below the blocking temperature, the magnetization fluctuations near the anisotropy energy minimum can be described using Boltzmann statistics \cite{morup2010magnetic}.
If  the anisotropy energy is expressed using only the first terms of the Taylor series, the temperature dependence of the hyperfine field is $B_\mathrm{obs} = B(T = 0\mathrm{K} )[1-k_\mathrm{B}T/(\kappa V)]$  where $\kappa$ is a parameter describing the curvature of the anisotropy near its energy minimum \cite{morup1983magnetic}.
For non-interacting cores with uniaxial anisotropy,  the energy is given as $E(\theta)=KV\sin^2(\theta)$ and consequently $\kappa V=2KV$. 
Fig.\,\ref{Fig3}(c)  shows the mean hyperfine field \textit{vs.} temperature. 
Using this method, we find $\kappa V$ to be basically indistinguishable for the two samples (powder: $\kappa V/k_\mathrm{B}=2200(100)$\,K, frozen dispersion: $2140(220)$\,K). 

If we assume a value of $K$ of $\mathrm{13\,kJ/m^3}$ (i.e. the upper limit reported in literature \cite{svedlindh1997intra}), then we expect $\kappa V/k_\mathrm{B}$ to be around 1000\,K for isolated 10-nm cores.  
The larger $\kappa V/k_\mathrm{B}$ values indeed indicate an additional contribution of the anisotropy for both systems, probably caused by intercore interactions.
In literature, anisotropy values larger than $\mathrm{13\,kJ/m^3}$ have been reported for maghemite particles, either due to interparticle interactions \cite{morup2010magnetic}, or due to increased surface spin disorder for smaller particles \cite{de2017remanence}.
But it can be noted that the broad asymmetric lines in the M\"ossbauer spectra observed at 200\,K (Fig.\,\ref{Fig3}(a) and (b)) are  typical for magnetic fluctuations governed by an interaction field arising from interparticle interactions \cite{fock2018interpretation}.
Additionally we can conclude from the observed splitting of the spectra at 200\,K that the cores are still thermally blocked, although it is worth mentioning that at 295(5)\,K they are clearly superparamagnetic on the M\"ossbauer time scale of $\approx1\,\mathrm{ns}$ (Fig.\,\ref{Fig3}(d)).
Furthermore, whilst we do not measure a discernible difference in $\kappa V$ for the two methods of sample preparation, at 200\,K it is clear that the spectrum of the particles in the frozen dispersion is slightly more collapsed than that of the powder.  
This points towards slightly weaker interactions between cores within the colloid. 

Fig.\,\ref{Fig4}(a) shows the normalized $M(H)/M_\mathrm{S}$ curves of the dispersion and powder measured at 300\,K. 
The saturation magnetization $M_\mathrm{S}$ was measured at $\mu_0H=5\,\mathrm{T}$, which for the powder was $67\,\mathrm{Am^2/kg}$ (normalized to the total sample mass).
Using the density $\rho=4869\,\mathrm{kg/m^3}$ from neutron diffraction (Table\,\ref{Table A}), which is close to the density of pure maghemite ($\rho=4860\,\mathrm{kg/m^3}$) \cite{coey}, this yields a volume saturation magnetization of $327\,\mathrm{kA/m}$, which is in good agreement with our observation using neutron diffraction.
The complete magnetization curves for both the colloidal dispersion and powder were anhysteretic, which indicates a superparamagnetic behavior on the measurement time scale $\approx$ 100\,s for the two samples due to thermal moment fluctuations.
Notably, the magnetic susceptibility of the powder is smaller at low and intermediate fields ($<1\,\mathrm{T}$) compared to that of the dispersion, which we attribute to a larger dipolar interaction field within the powder.

Clear signatures of dipolar interactions in the powder are also observed when comparing the in-phase ($\chi'$) and out-of-phase components ($\chi''$) of the ACS($T$) susceptibilities of the two systems (Fig.\,\ref{Fig4}(b)). 
$\chi'$ and $\chi''$ of the powder are suppressed, broadened and shifted towards higher temperatures relative to the dispersion. 
In both samples, $\chi''=0$ for $T>250\,\mathrm{K}$, and which is indicative of superparamagnetic behavior on the studied time scales (characteristic measurement times in the range $10^{-3}-1\,\mathrm{s}$).
We have deduced the frequency dependence of the blocking temperature, $T_\mathrm{B}(f)$, whereby $T_\mathrm{B}$ was defined as the temperature for which $\chi''$ has reached 50\,\% of its maximum value.
Rather interestingly, if we use the N\'eel-Arrhenius equation ($\tau=\tau_0 \cdot\mathrm{exp}(-KV/k_\mathrm{B}T)$) to fit our data then we obtain grossly unphysical values of $\tau_0$, of the order of $10^{-40}\,\mathrm{s}$, for both the frozen dispersion as well as the powder.  
This in and of itself, is a clear indication of the presence of significant intercore interactions. 
At this point, it might be tempting to define the critical slowing down process, which may be taking place here, and eventually disclose the intimate nature of the spin dynamics relaxation. 
It is very likely that the behavior could be related to a super-spin glass state at low temperature \cite{alonso2010crossover}.
However, for this study we focus in the following on the analysis of the ACS($T$) data in the superparamagnetic regime.
We surmise, that the observed interactions arise from the dense agglomeration of the cores to clusters, which we revealed by SAXS.  
To further investigate the local dipolar interaction field in both systems, we fitted $\chi'$ in the superparamagnetic regime (where $\chi''$ = 0) to a mean-field model:
\begin{align}\label{Eq2}
\chi^{-1}=\frac{T-\alpha C(0) (1-BT^{\beta})^2}{C(0)(1-BT^{\beta})^2},
\end{align} 
with $C(0)=\mu_0V_\mathrm{p}/(3k_\mathrm{B})\cdot M_\mathrm{S}^2(0)$.
Here, the dipolar field $H_\mathrm{d}=\alpha M$ is given by the mean field constant $\alpha$ and the field-induced magnetization, where $\alpha>0$ is indicative of a (on average) FM-like coupling, $\alpha<0$ of an AFM-like coupling, and $\alpha=0$ for a non-interacting system \cite{faure20132d}.
The fitting procedure of the inverse susceptibility ($1/\chi$) using the Curie-Weiss mean-field model was then as follows.
First, we determined the temperature dependency of the magnetization.
By fitting the $M(T)$ curve measured in a field of $\mu_0H=1\,\mathrm{T}$ to a modified Bloch-law: $M_\mathrm{S}(T)=M_\mathrm{S}(0) (1-BT^{\beta})$ (Fig.\,\ref{Fig4}(c)). 
Omitting the slight upturn in the low temperature interval ($5-40\,\mathrm{K}$), this yields $\beta= 2.18$ and a corresponding value of $B=4.25 \times 10^{-7} K^{-2.18}$. 
In the second step, we corrected the data for a small deviation between the $M_\mathrm{S}$ (at 5\,T) of the dispersion and the powder, using the absolute magnetization value of the powder as the reference ($66.5\,\mathrm{Am^2/kg}$ at 300\,K and 5\,T). 
Lastly, the inverse of the equilibrium (frequency-independent) part of the in-phase component of the magnetic susceptibility  ($1/\chi'$) was fitted to the Curie-Weiss law (Eq.\,\ref{Eq2}), using $\alpha$ and $C(0)$ as fit parameters.
These fits are shown in Fig.\,\ref{Fig4}(d). 
The differences between the $C(0)$ parameter derived for the powder (2800(10)\,K) and the frozen dispersion (2750(20)\,K) are small. 
Using the determined saturation magnetization for $T=0\,\mathrm{K}$ ($M_S(0)=362\,\mathrm{kA/m}$, Fig.\,\ref{Fig4}(c)) yields a core diameter of  11\,nm, which is in good agreement with the effective core diameters determined by TEM.
Examination of the coupling parameter $\alpha$, indicates a value of $\alpha = -0.0137(6)$ for the frozen dispersion and $\alpha = -0.0737(5)$ for the powder.
This difference indicates increased dipolar interactions for the powder compared to the frozen dispersion, as also observed by M\"ossbauer spectroscopy.
Nonetheless, the negative $\alpha$ values allow us to conclude that both systems have preferential AFM-like coupling between neighboring core moments in the clusters, which is irrespective of the sample preparation.

Summarizing this section, our pre-characterization revealed the existence of a major maghemite core contribution with a ferrimagnetic spin arrangement.
The core sizes lie around 10\,nm but they are agglomerated to clusters of about 70\,nm.
Magnetometry reveals a superparamagnetic behavior at 300\,K, but which is accompanied by interparticle dipolar interactions, and which seem to result in preferentially an AFM-like moment coupling.

\subsection{Analysis of the POLARIS experiment}

\begin{figure}[t]
\centering
\includegraphics[width=1\columnwidth]{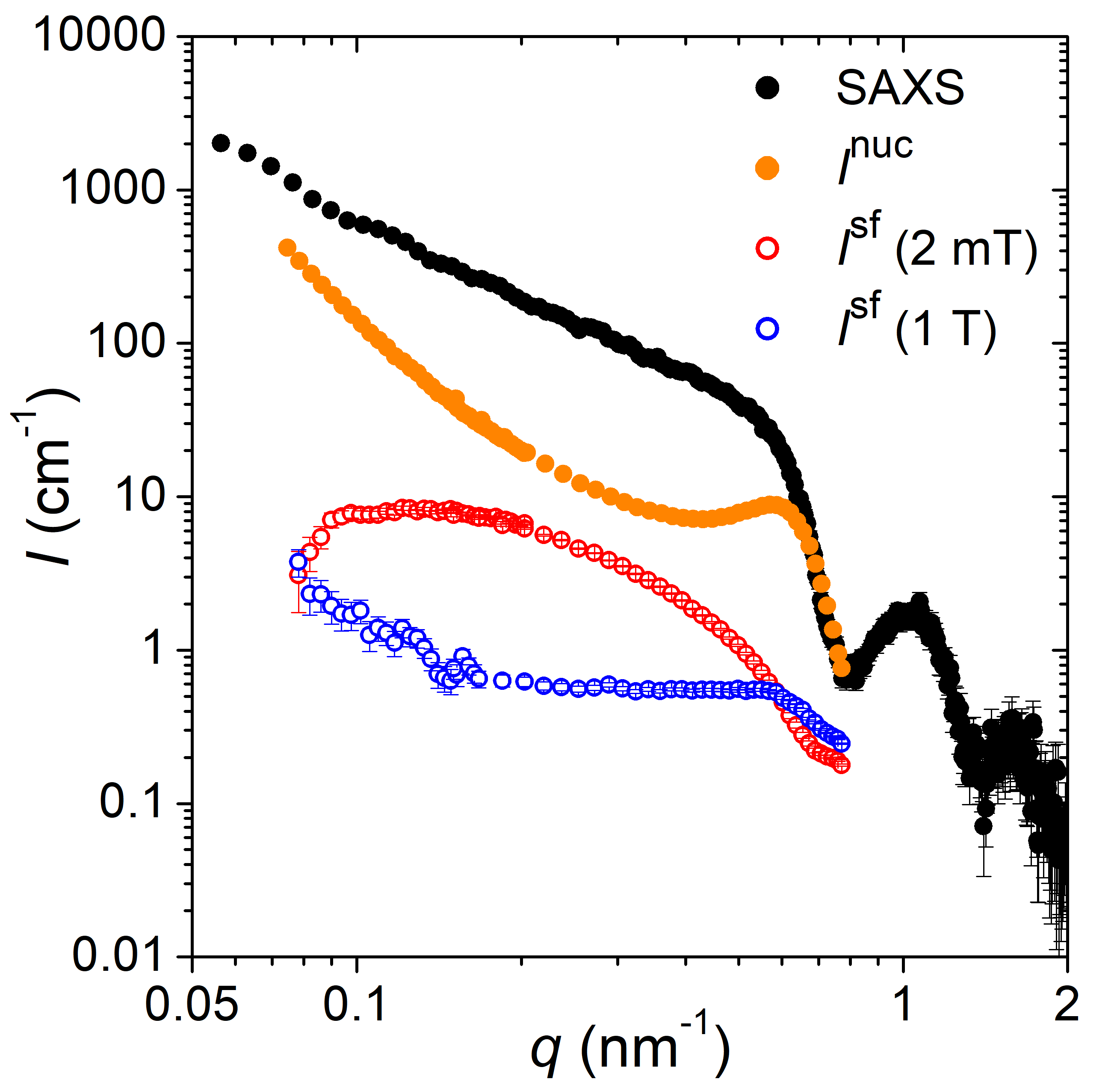}
\caption{\label{Fig5}
SAXS intensity from Fig.\,\ref{Fig1} (colloidal dispersion) and the nuclear SANS intensity $I^{\mathrm{nuc}}(q)$ (powder; rescaled), derived from the non-spin-flip intensities of the POLARIS experiment. Additionally we show here the radial average of the spin-flip Intensities $I^{\mathrm{sf}}(q)$ measured at 2\,mT and 1\,T (same as in \textit{inset} of Fig.\,\ref{Fig6}(c); in arbitrary units).}
\end{figure}

Fig.\,\ref{Fig5} shows the purely nuclear SANS cross section $I^{\mathrm{nuc}}(q)$, derived from the POLARIS experiment, as described in the appendix.
Similar to the SAXS data of the colloid, the nuclear SANS data of the powder exhibits an increasing intensity for decreasing $q$, which can be attributed to the fractal structure of the core-clusters.
In contrast to SAXS, however, $I^{\mathrm{nuc}}(q)$ contains a well-pronounced peak in the medium $q$-range.
The position in reciprocal space $q=0.58\,\mathrm{nm^{-1}}$ corresponds to a real-space size of around $2\pi/q=11\,\mathrm{nm}$.
This correlates well to the expected core-to-core distance between neighboring particles within the clusters and thus the peak can be attributed to the interparticle correlations (i.e. structure factor).
The fact that this correlation peak is well visible in SANS but not in SAXS indicates on average a closer packing of the aggregates in powder.
This can be simply explained by the evaporation of the water due to the freeze-drying process, which results in a collapse of the polymer shell compared to the swollen state in colloidal dispersion.

\begin{figure*}[ht]
\centering
\includegraphics[width=1\textwidth]{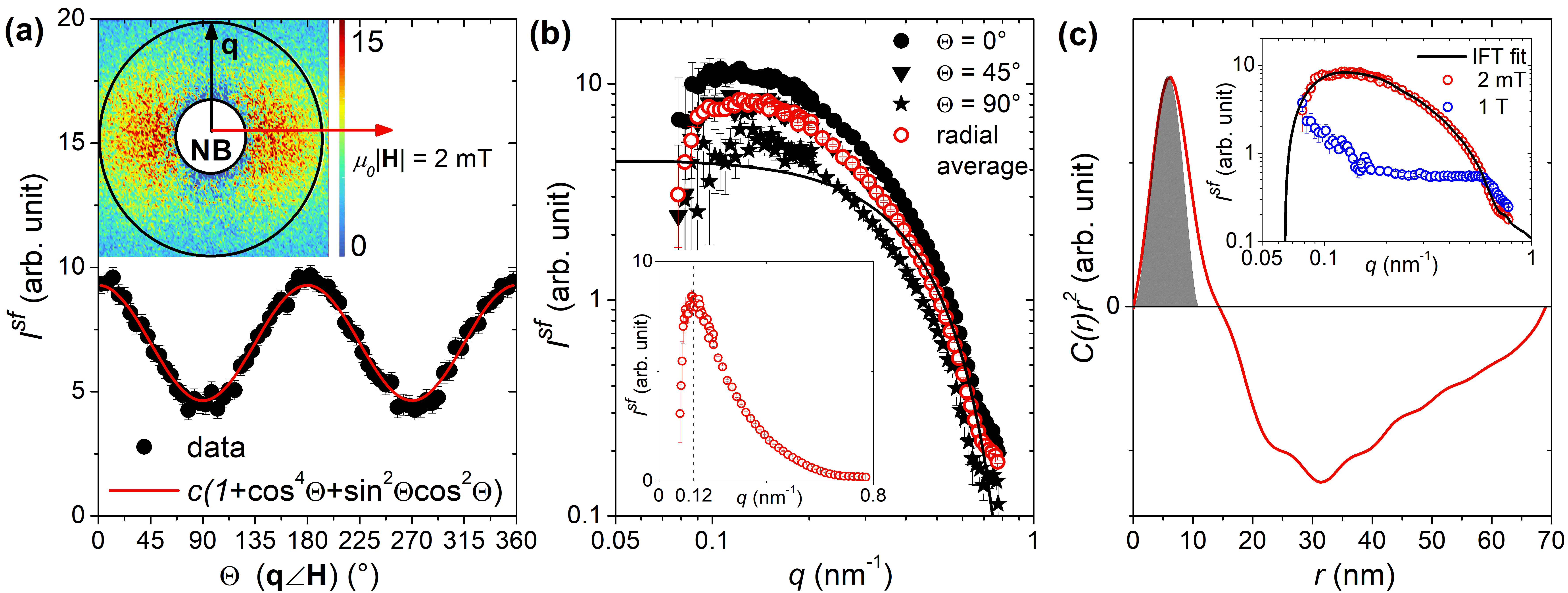}
\caption{\label{Fig6}Results of the polarized SANS experiments of the powder at 300\,K:
\textbf{(a)} Spin flip intensity $I^{\mathrm{sf}}(\textbf{q})$ ($\mu_0H=2\,\mathrm{mT}$) integrated over $|\textbf{q}|=0.07-0.21\,\mathrm{nm^{-1}}$ as a function of $\Theta$. \textit{inset}: Corresponding 2D scattering pattern $I^{\mathrm{sf}}(\textbf{q})$ (detector distance 13.4\,m). NB indicates area of the primary neutron beam.
\textbf{(b)} Azimuthal averages of $I^{\mathrm{sf}}(\textbf{q})$ ($\mu_0H=2\,\mathrm{mT}$) in $10^\circ$ sectors around $\Theta=0^\circ$, $45^\circ$, $90^\circ$ and the radial average, as well as the form factor $F(q)$ of a sphere (Eq.\,\ref{EqA3}).
\textbf{(c)} The correlation function $C(r)r^2$ determined by an IFT of the radial average of $I^{\mathrm{sf}}(\textbf{q})$ measured at 2\,mT, and the profile of a homogeneous sphere calculated with Eq.\,\ref{Eq1} for $D=10\,\mathrm{nm}$ (grey area). 
\textit{inset}: Comparison of the radial average $I^{\mathrm{sf}}(q)$ measured at 2\,mT and 1\,T, and fit of the measurement at 2\,mT by IFT.}
\end{figure*}

The purely magnetic scattering is presented in Fig.\,\ref{Fig6}.
The \textit{inset} of Fig.\,\ref{Fig6}(a) displays the 2D scattering pattern $I^{\mathrm{sf}}(\textbf{q})$ detected for the detector distance 13.4\,m at 300\,K and a field strength $\mu_0H$ of 2\,mT.
At the same field strength, isothermal magnetization measurements reveal that the normalized magnetization is a very small value of 0.026 (Fig.\,\ref{Fig4}(a)), and which suggests that, on average, the moments within the ensemble were basically randomly distributed.  
We confirm this by the angular dependency of $I^{\mathrm{sf}}(\textbf{q})$ integrated over $|\textbf{q}|=0.07-0.21\,\mathrm{nm^{-1}}$.
It is clear that $I^{\mathrm{sf}}(\textbf{q})$ obeys a $(1+\mathrm{cos}^4\Theta+\mathrm{sin}^2\Theta\mathrm{cos}^2\Theta$) behavior (Fig.\,\ref{Fig6}(a)), and which we expect for magnetization equal in $x$, $y$ and $z$-direction with an isotropic distribution of moments around the $y$ and $z$-axis ($\widetilde{M}_y\widetilde{M}_z^*+\widetilde{M}_z\widetilde{M}_y^*=0$, Eq.\,\ref{EqA2}).
Note that for the case of magnetically non-interacting cores in zero field, the scattering cross section is proportional to the single-particle form factor \cite{wiedenmann2005polarized}.

The azimuthal averages of $I^{\mathrm{sf}}(\textbf{q})$ in $10^\circ$ sectors around $\Theta=0^\circ$, $45^\circ$ and $90^\circ$ are plotted in Fig.\,\ref{Fig6}(b), and which depend on the superposition of the individual cross sections $|\widetilde{M}_x|^2$,$|\widetilde{M}_y|^2$ and $|\widetilde{M}_z|^2$ (Eq.\,\ref{EqA2}).
The absolute values of the intensity decrease from $0^\circ$ to $45^\circ$ to $90^\circ$, but the functional form is basically identical.
This further strengthens our hypothesis that a small applied field is not enough to result in significant alignment of the moments in the direction of the field, otherwise the shape of $|\widetilde{M}_z|^2$ would strongly deviate from that of $|\widetilde{M}_{x,y}|^2$.
Due to the observed isotropy of the magnetization, we can focus our analysis on the radial average $I^{\mathrm{sf}}(q)=\int_0^{2\pi}I^{\mathrm{sf}}(\textbf{q})\mathrm{d}\Theta\approx3/2|\widetilde{\mathbf{M}}|^2$ (Fig.\,\ref{Fig6}(b)).

A characteristic feature of $I^{\mathrm{sf}}(q)$ is its maximum at $0.12\,\mathrm{nm^{-1}}$ and decrease with decreasing $q$ (Fig.\,\ref{Fig6}(b)).
This peak becomes in particular visible in linear scale as shown in the \textit{inset} of Fig.\,\ref{Fig6}(b).
The decrease is in contrast to the form factor $F(q)$ of a single sphere (Eq.\,\ref{EqA3}), which monotonically increases in the Guinier regime ($q<0.26\,\mathrm{nm^{-1}}$ for 10-nm spheres; black line in Fig.\,\ref{Fig6}(b)).
Similar peaks have been also observed in other studies \cite{sankar2000magnetic,avdeev2004magnetic,farrell2006small,sachan2008field}.
One would expect to observe this form factor if the cores were indeed not magnetically interacting; the fact that we do not observe this shows that there are magnetic interactions present. 

This can be further verified by the underlying magnetic correlation function (Fig.\,\ref{Fig6}(c)), which we derived from $I^{\mathrm{sf}}(q)$ by an IFT \cite{bender2017structural}, as outlined in the \hyperref[appendix]{appendix}.
The maximal size $D_{\mathrm{max}}$ according to the correlation function (i.e. where the $C(r)$ reaches zero) is 69\,nm.
The derived correlation function fits the experimental data well within the accessible $q$-range (\textit{inset} of Fig.\,\ref{Fig6}(c)), and thus we can assume that $C(r)$ correctly represents the moment correlations, at least in the low $r$-range (i.e. between nearest and next-nearest neighbors).
As can be seen, $C(r)r^2$ is positive for $r<15\,\mathrm{nm}$, but takes on negative values for $15\,\mathrm{nm}<r<69\,\mathrm{nm}$.

The primary peak of $C(r)r^2$ is well described by the calculated profile of a homogeneous sphere with a diameter of 10\,nm (Eq.\,\ref{Eq1}).
As a reminder, the Fourier transform of the correlation function Eq.\,\ref{Eq1} is just the single-particle form factor plotted in Fig.\,\ref{Fig6}(b) (black line; see also Eq.\,\ref{EqA3}).
Hence, this profile corresponds to the expected magnetic correlation function of the (isolated) homogeneously magnetized cores and confirms their single-domain state.
The $C(r)$ function for $r>10\,\mathrm{nm}$, then, describes the intercore moment correlations.
The fact that $C(r)$ crosses zero at $r=15\,\mathrm{nm}$ suggests for neighboring particle moments on average a competition between positive (i.e. FM-like alignment) and negative moment correlations (i.e. AFM-like alignment).
The negative values of the correlation function for $r>15\,\mathrm{nm}$, however, verify that on average the core moments of next-nearest neighbors tend to align antiparallel to each other. 

\begin{figure*}[t]
\centering
\includegraphics[width=1\textwidth]{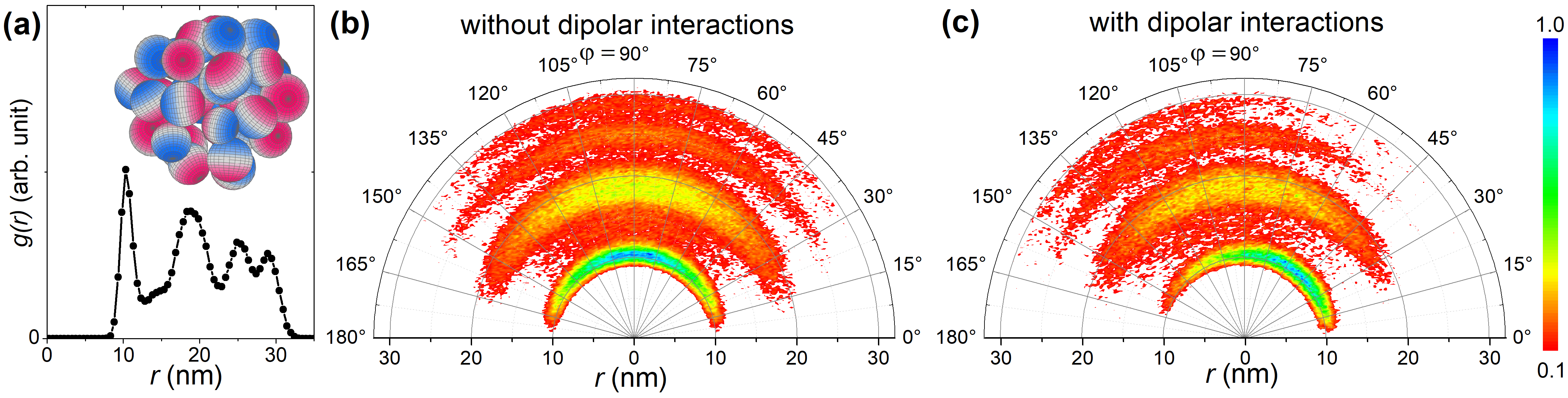}
\caption{\label{Fig7}Monte Carlo simulations: 
(a) Averaged radial distribution function $g(r)$ (160 bins) of the ensemble of 320 clusters with 32 cores each. \textit{inset}: \textit{Snapshot} of one simulated cluster. 
The volume of the spheres is directly proportional to the magnetic moment, and red-blue caps indicate the direction of the anisotropy axis.
(b) Normalized polar plot of the magnetic pair correlations within the clusters without dipolar interactions and (c) with dipolar interactions.}
\end{figure*}

In the \textit{inset} of Fig.\,\ref{Fig6}(c) we have additionally plotted the radial average of the spin-flip SANS intensity detected at 1\,T.
Here the scattering intensity monotonically increases with decreasing $q$, similar to the purely nuclear scattering intensity $I^{\mathrm{nuc}}(q)$ (Fig.\,\ref{Fig5}).
This suggests strictly positive correlations between neighboring core moments at this field strength, which can be simply explained by a parallel alignment of the core moments along the applied magnetic field ($z$-axis).

\subsection{Monte Carlo simulations}

By SAXS we saw that the 10-nm cores were agglomerated to clusters with sizes of around 68\,nm.
Analysis of the AC susceptibility measurements of the frozen dispersion in the superparamagnetic regime then indicated a preferential AFM-like coupling between the core moments within the clusters.
Furthermore, we could show that, by compacting the core-clusters to a powder, the coupling strength was increased, and polarized SANS verified on average an AFM-like coupling between next-nearest neighbor moments.
Thus to confirm the possibility for anticorrelations between fluctuating core moments, we have simulated the magnetic properties such core-clusters.
In the literature, a number of different approaches are used to theoretically reveal the influence of dipolar interactions on the magnetization behavior of magnetic core-clusters \cite{tan2014magnetic,ovejero2016effects,anand2016spin,usov2017interaction,schaller2009monte,schaller2009effective,melenev2010monte,hovorka2014role,ilg2017equilibrium,kure2017magnetic}.
In the current work, we have used Monte Carlo simulations and focused on the determination of the directional correlations between the core moments.

For the Monte Carlo simulations, we used a constant pressure approach to generate an ensemble of 320 clusters with each cluster containing 32 individual cores \cite{miller2015dynamics}.
Each core was modeled as a point dipole with a spherical exclusion volume that was proportional to the magnetic moment and had an anisotropy energy density of $13\,\mathrm{kJ/m^3}$. 
We also ensured that the corresponding size distribution closely resembled the one observed by TEM.
To minimize the interaction energy of the cores, both the orientation and position were sampled within the simulation. 
Fig.\,\ref{Fig7}(a) displays the average radial distribution function (i.e. pair correlation function $g(r)$) of the clusters, and the \textit{inset} of Fig\,\ref{Fig7}(a) shows the realization of such a cluster. 
It is immediately clear that the distribution of the clusters shows 4 distinct correlation peaks that indicate well ordered structures. 

By using kinetic Monte Carlo similations \cite{tan2014magnetic}, we have been able to access information into the magnetization dynamics of the core-clusters at 293\,K and at an applied field $\mu_0H$ of 2\,mT.
For each of the 320 clusters we determined for all 496 unique core-pairs the magnetic pair correlation function $g_{ij}(t)=\mathbf{m}_i(t)\cdot\mathbf{m}_j(t)=m_im_j\mathrm{cos}\varphi(t)$ at $1000$ different time points.
The $496\times320\times1000$ unique $g_{ij}$ functions were binned in $601\times 601$  pairs of $\lbrace r,\varphi\rbrace$, with $r=|\mathbf{r}_{ij}|$ being the distance between the core centers.
Fig.\,\ref{Fig7}(b) shows the resulting 3D polar plot $\left[r,\varphi,P(r,\varphi)\right]$, with $P(r,\varphi)$ being the sum of all moment products $m_im_j$ for each bin (here we normalized $P$ to the global maximum).
This means, the $P(r,\varphi)$ can be regarded as proportional to the probability that the angle between two moments displaced by $r$ amounts to $\varphi$.

For the simulated clusters without dipolar interactions we observe that at a given $r$-value the probability $P(\varphi)$ displays a $\mathrm{sin}\varphi$-like dependence (Fig.\,\ref{Fig7}(b)), which is the expected time average for an isotropic ensemble of moments, whose dynamics is only governed by the randomly distributed anisotropy axes of the individual cores.

However, when we include dipolar interactions of the form
\begin{equation}\label{Eq3}
    E_{\mathrm{dip},i}=\frac{-\mu_0}{4\pi}\sum_{j\neq i}\frac{\big(3(\mathbf{m_i}\cdot \mathbf{\hat{r}}_{ij})(\mathbf{m_j}\cdot \mathbf{\hat{r}}_{ij})-(\mathbf{m_i}\cdot \mathbf{m_j})\big)}{r_{ij}^3},
\end{equation}
the distribution of magnetic pair correlations gets distorted (Fig.\,\ref{Fig7}).
At $r\sim10\,\mathrm{nm}$ (i.e. nearest neighbor regime) the maximum of $P(\varphi)$ is shifted to $\varphi<90^\circ$, which clearly indicates a tendency to rotate neighboring moments parallel to each other.
For larger $r$-values, however, the maximums of $P(\varphi)$ shifts to $\varphi>90^\circ$, which means that the dipolar interactions induce for next-nearest neighbor moments an inclination towards an AFM-like alignment, thus confirming our previous experimental findings.

\section{Conclusions}

In this study we have investigated the moment coupling between iron oxide cores, which were agglomerated to clusters, and found strong evidence for directional correlations between neighboring core moments also in the superparamagnetic regime.
According to TEM and neutron diffraction the cores have a mean diameter of around 10\,nm.
The ensemble properties were then analyzed both in (frozen) colloidal dispersion and in powder form.
Analysis of the SAXS intensity of the colloid revealed that the as-prepared cores were agglomerated to clusters with sizes of around 68\,nm, which agrees well with DLS. 
A combination of temperature-dependent M\"ossbauer spectra and the Rietveld refinement of a neutron diffraction pattern shows that the cores were composed of $>95\,\%$ maghemite, $<5\,\%$ magnetite.
Additionally, the analysis of the M\"ossbauer spectra indicated strong dipolar interactions between the core moments within the core-clusters in both the liquid dispersion and in the powder. 
The magnetization measurements of the dispersion and powder showed that at 300\,K the particles behaved macroscopically superparamagnetic despite clear signs of dipolar interactions.
Analysis of temperature-dependent AC susceptibility data implied dipolar coupled anticorrelations between the thermally fluctuating core moments in both systems.
To further reveal the nature of the coupling we performed a polarized SANS experiment on the powder: by applying POLARIS we detected the purely magnetic cross sections at 300\,K and at an applied field $\mu_0H$ of 2\,mT, from which we extracted the underlying magnetic correlation function by an indirect Fourier transform.
For nearest neighbors the extracted distribution indicated a competition between an FM-like and an AFM-like coupling.
This tendency was also found by kinetic Monte Carlo simulations of such core-clusters.
For moments located further away, however, the simulations exhibited an inclination towards an AFM-like alignment.
This is in good agreement with our polarized SANS experiment, where the derived distribution function clearly verified a preference for anticorrelations between next-nearest core moments.

\begin{acknowledgments}
We thank the Institut Laue Langevin for provision of beamtime at the instruments D2B and D33.
This project has received funding from the European Commission Framework Programme 7 under grant agreement no 604448 (NanoMag).
C.F. also acknowledges funding from the Independent Research Fund Denmark.
\end{acknowledgments}

\newpage

\section*{Appendix}\label{appendix}
\setcounter{table}{0}
\setcounter{equation}{0}

\renewcommand{\thetable}{A\arabic{table}}
\renewcommand{\theequation}{A\arabic{equation}}

\begin{table}[h]
\caption{Results from the Rietveld refinement of the neutron diffraction pattern, for the cubic $Fd\bar{3}m$ inverse spinel space group at 300\,K [Fe tetrahedral site at $(1/8, 1/8, 1/8)$; Fe Octahedral site at $(1/2, 1/2, 1/2)$; O at $(1/4 + u, 1/4 + u, 1/4 + u)$].
Lattice parameter $a$, O coordinate $u$, isotropic thermal parameter $B_\mathrm{iso}$, density $\rho$, occupancy of the Fe octahedral site Occ, magnetic moment $\mu$ at tetrahedral and octahedral sites, average crystal/magnetic size $D_\mathrm{cryst/mag}$, crystal/magnetic microstrain $\epsilon_\mathrm{cryst/mag}$, as well as agreement factors $R_\mathrm{p}$, $R_\mathrm{wp}$, $R_{\mathrm{B}}$, $R_{\mathrm{mag}}$ and the goodness of fit, $\chi^2$.}
\label{Table A}
\begin{tabular}{l l c}
parameters & & results \\
\hline\hline
$a$ (\AA)                & & 8.3565(3) \\ 
$u$                                 & & 0.0059(2) \\
$B_\mathrm{iso}$ $\mathrm{(\AA^2)}$ Fe-tet & & 0.86(8) \\
\ \ \ \ \ \ \ \ \ \ \ \ \ Fe-oct                                 & & 1.0(1) \\
\ \ \ \ \ \ \ \ \ \ \ \ \ O                                   &  & 0.09(6) \\
$\rho$ $\mathrm{(kg/m^3)}$          & & 4869(1) \\
Occ  \footnote{Occupancy was fixed for the refinement and estimated from the result of M\"ossbauer spectroscopy ($95/5\,\mathrm{wt\%}$  $\gamma$-$\mathrm{Fe_2O_3}$/$\mathrm{Fe_3O_4}$). Expected values for pure maghemite and magnetite are 1 and $5/6$, respectively.}                               & & 0.84167 \\
$\mu/\mu_\mathrm{B}$ Fe-tet                     & & 4.5(2) \\
\ \ \ \ \ \ \ \,\,\,Fe-oct                                 &  & 4.2(2) \\
$D_\mathrm{cryst}$ (nm)                  & & 9(1) \\
$\epsilon_\mathrm{cryst}$ (\textpertenthousand)        & & 29(6) \\
$D_\mathrm{mag}$ (nm)                 & & 6(2) \\
$\epsilon_\mathrm{mag}$ (\textpertenthousand)   \footnote{Magnetic strain was assumed to be the same as crystal strain.}      & & 29(6) \\
$R_\mathrm{p}$ (\%)                          & & 2.52 \\
$R_\mathrm{wp}$ (\%)                       & & 3.16 \\
$R_{\mathrm{B}}$ (\%)                    & & 5.77 \\
$R_{\mathrm{mag}}$ (\%)                      & & 9.75 \\
$\chi^2$                        & & 1.39 
\end{tabular}
\end{table}

\FloatBarrier
\newpage
\subsection*{Analysis of the polarized SANS data}

To separate magnetic from nuclear scattering contributions, we performed SANS with POLARIS option \cite{honecker2010longitudinal,michels2014magnetic}.
The purely nuclear SANS cross section can be extracted from the non-spin-flip cross sections $I^{++}(\textbf{q})$, $I^{--}(\textbf{q})$ (for $\mathbf{H}\perp \mathbf{k}$):
\begin{align}\label{EqA1}
I^{\pm\pm}(\mathbf{q})\propto&|\widetilde{N}|^2+b_\mathrm{h}^2|\widetilde{M}_z|^2\mathrm{sin}^4\Theta\nonumber\\
&+b_\mathrm{h}^2|\widetilde{M}_y|^2\mathrm{sin}^2\Theta\mathrm{cos}^2\Theta\nonumber\\
&-b_\mathrm{h}^2(\widetilde{M}_y\widetilde{M}_z^*+\widetilde{M}_z\widetilde{M}_y^*)\mathrm{sin}^3\Theta\mathrm{cos}\Theta\nonumber\\
&\mp b_\mathrm{h}(\widetilde{N}\widetilde{M}_z^*+\widetilde{N}^*\widetilde{M}_z)\mathrm{sin}^2\Theta\nonumber\\
&\pm b_\mathrm{h}(\widetilde{N}\widetilde{M}_y^*+\widetilde{N}^*\widetilde{M}_y)\mathrm{sin}\Theta\mathrm{cos}\Theta.
\end{align}
Here $\Theta$ is the angle between the scattering vector $\mathbf{q}=(0,q_y,q_z)$ and the magnetic field $\mathbf{H}$ and $b_\mathrm{h}=2.7\cdot10^{-15}\,\mathrm{m}/\mu_\mathrm{B}$, where $\mu_\mathrm{B}$ is the Bohr magneton. 
Moreover, $\tilde{N}(\vec{q})$ and $\widetilde{\mathbf{M}}=\left[\widetilde{M}_x(\textbf{q}),\widetilde{M}_y(\textbf{q}),\widetilde{M}_z(\textbf{q})\right]$ denote the Fourier transforms of the nuclear scattering length density and of the magnetization in the $x$-, $y$- and $z$-directions, respectively, and the index $^*$ the complex conjugate.
Hence, the purely nuclear cross section $I^{\mathrm{nuc}}(q)\propto|\tilde{N}|^2$ (here assuming isotropy), on the one hand, can be determined from the sector parallel to $\mathbf{H}$ of the non-spin-flip intensities.
The spin-flip intensities, on the other hand, are of purely magnetic origin.
For our sample we assume that chiral scattering terms can, at first approximation, be neglected \cite{michels2014magnetic}, and thus we can write $I^{\mathrm{sf}}(\textbf{q})=I^{+-}(\textbf{q})=I^{-+}(\textbf{q})$, with\cite{honecker2010longitudinal}
\begin{align}\label{EqA2}
I^{\mathrm{sf}}(\mathbf{q})\propto&|\widetilde{M}_x|^2+|\widetilde{M}_y|^2\mathrm{cos}^4\Theta+|\widetilde{M}_z|^2\mathrm{sin}^2\Theta\mathrm{cos}^2\Theta\nonumber\\
&-(\widetilde{M}_y\widetilde{M}_z^*+\widetilde{M}_z\widetilde{M}_y^*)\mathrm{sin}\Theta\mathrm{cos}^3\Theta.
\end{align} 
for $\mathbf{H}\perp \mathbf{k}$.
For homogeneously magnetized and non interacting nanoparticles (i.e. single-domain, single-core), the functional forms of $\left[\widetilde{M}_x(\textbf{q}),\widetilde{M}_y(\textbf{q}),\widetilde{M}_z(\textbf{q})\right]$ are considered to be proportional to the single-particle form factor, $F(\textbf{q})$.
For a spherical particle with radius $R$ 
\begin{align}\label{EqA3}
F(q)&=\left[\frac{3}{qR}\left(\frac{\mathrm{sin}(qR)}{(qR)^2}-\frac{\mathrm{cos}(qR)}{qR}\right)\right]^2\\
&=\frac{1}{R}\int_0^{2R}C(r)r^2\frac{\mathrm{sin}(qr)}{qr}\mathrm{d}r,\nonumber
\end{align}
with $C(r)$ being the correlation function from Eq.\,\ref{Eq1} \cite{svergun2003small}.

It is possible to obtain the real-space correlation function $C(r)$ by a Fourier transform of the reciprocal scattering data, which is a model-free description of the underlying structure giving rise to small-angle scattering \cite{stuhrmann1970interpretation,glatter1977new}.
In \citet{michels2003range} for example a direct Fourier transform was applied to derive $C(r)$ from magnetic SANS and, by analyzing the extracted functions, enabled the authors to determine characteristic magnetic correlation lengths within crystalline soft nanomagnets. 
In case of nuclear scattering, the usual approach is to apply an indirect Fourier transform (IFT) of the scattering intensities to extract the autocorrelation function $C(r)$ (or $C(r)r^2$ to emphasize long-range correlations) of the nuclear scattering length density, as also performed on the SAXS data in this manuscript (Fig.\,\ref{Fig1}(b)).
We then applied the same approach to derive the underlying magnetic correlation function $C(r)r^2$ from the purely magnetic spin-flip SANS data of the particulated system of macrospins.

For the IFT we used the procedure described in \citet{bender2017structural} where the maximal size $D_\mathrm{max}$ (i.e. the distance $r$ for which $C(r>D_{\mathrm{max}})=0$) of the correlation function is a free fit parameter.
This parameter can be usually estimated from the low $q$ scattering behavior in the Guinier regime.
In our case, however, the Guinier regime is not reached ($I^{\mathrm{sf}}(q)$ is expected to approach a constant value $>0$ for $q\rightarrow 0$), and thus we could not derive the complete correlation function, but only an estimation in the nearest neighbor range.
Our approach was then as follows: we varied $D_\mathrm{max}$ in 1-nm steps from $10-100\,\mathrm{nm}$ ($q=0.07\,\mathrm{nm^{-1}}$ corresponds to a real space size of $r=2\pi/q\approx 90\,\mathrm{nm}$), performed for each $D_\mathrm{max}$ value the IFT to determine the corresponding correlation function with 100 bins and calculated subsequently the evidence by a Bayesian analysis \cite{hansen2000bayesian}.
In Fig.\,\ref{Fig6}(c) we plotted the function $C(r)r^2$ for which the largest evidence was calculated.

Regarding the interpretation of $C(r)$ it has to be considered that the extracted correlation function
\begin{equation}\label{EqA4}
C(r)\propto \int I^{\mathrm{sf}}(\mathbf{q}) \mathrm{exp}\left( i \mathbf{q}\mathbf{r}\right)\mathrm{d}\mathbf{q}
\end{equation}
from magnetic SANS data (which are folded with the magnetodipolar interaction of the neutron, entering \textit{via} the trigonometric functions of $\Theta$ in Eq.\,\ref{EqA2}) are not necessarily the autocorrelation function of the magnetization vector field
\begin{align}\label{EqA5}
C_A(r) &\propto \int \textbf{M}(\textbf{x})\textbf{M}(\textbf{x}+\textbf{r})\mathrm{d}\mathbf{x}\\
&\propto\int\left[ |\widetilde{M}_x|^2 + |\widetilde{M}_y|^2 + |\widetilde{M}_x|^2\right] \mathrm{exp}\left( i \mathbf{q}\mathbf{r}\right) \mathrm{d}\mathbf{q}\nonumber,
\end{align}
as discussed in \citet{mettus2015small} and \citet{erokhin2015dipolar}.
Yet in case of isotropy and equality of the cartesian magnetization components (which is at first approximation the case for low fields in our case) Eqs.\,\ref{EqA4} and \ref{EqA5} are qualitatively identical.
For homogeneously magnetized, spheres it can be thus assumed that without particle interactions the derived $C(r)$ equals Eq.\,\ref{Eq1}. 
An \textit{ad hoc} interpretation of the extracted correlation functions is then possible in so far, that positive values for $r>D$ (with $D$ being the core diameter) indicate on average a FM-like alignment and negative values an AFM-like alignment (anticorrelations) of the particle moments.
Anticorrelations due to dipolar stray fields were for example observed by \citet{erokhin2015dipolar} in case of inhomogeneous bulk ferromagnets, which manifested itself in negative values of the extracted correlation functions $C(r)$.

\bibliography{PBenderBib}

\end{document}